\documentclass[12pt]{article}
\UseRawInputEncoding
\usepackage{authblk}
\usepackage{graphicx}
\usepackage{geometry}
\usepackage{caption}
\newgeometry{vmargin={20mm}, hmargin={20mm,20mm}}   
\usepackage[english]{babel}
\usepackage{ragged2e}
\usepackage{amsmath}
\usepackage{amsfonts}
\usepackage{amssymb}
\usepackage{graphicx}
\usepackage{verbatim}
\usepackage[T1]{fontenc}
\usepackage{fancyhdr}

\providecommand{\keywords}[1]
{
	\small	
	\textbf{\textit{Keywords---}} #1
}

\title{Silicon Drift Detectors spectroscopic response during the SIDDHARTA-2 Kaonic Helium run at the DA$\Phi$NE collider}

\author[1]{Marco~Miliucci*} 
\author[1]{Massimiliano~Bazzi} 
\author[2]{Damir~Bosnar} 
\author[3]{Mario~Bragadireanu} 
\author[4]{Marco~Carminati}
\author[5]{Michael~Cargnelli}
\author[1]{Alberto~Clozza}
\author[1]{Catalina~Curceanu}
\author[4]{Griseld~Deda}
\author[1]{Luca~De~Paolis}
\author[6,1]{Raffaele~Del~Grande}
\author[4]{Carlo~Fiorini}
\author[1]{Carlo~Guaraldo}
\author[1]{Mihail~Iliescu}
\author[7]{Masahiko~Iwasaki}
\author[1]{Pietro~King}
\author[1]{Paolo~Levi~Sandri}
\author[4]{Johann~Marton}
\author[8]{Pawe{\l}~Moskal}
\author[1]{Fabrizio~Napolitano}
\author[8]{Szymon~Nied\'zwiecki}
\author[9]{Kristian~Piscicchia}
\author[1]{Alessandro~Scordo}
\author[1]{Francesco~Sgaramella*}
\author[1]{Hexi~Shi}
\author[8]{Micha{\l}~Silarski}
\author[1,3]{Diana~Sirghi}
\author[1,3]{Florin~Sirghi}
\author[8,1]{Magdalena~Skurzok}
\author[1]{Antonio~Spallone}
\author[5]{Marlene~T\"{u}chler}
\author[1,6]{Oton~Vazquez~Doce}
\author[5]{Johann~Zmeskal}

\affil[1]{\\INFN, Laboratori Nazionali di Frascati, Frascati, 00044 Roma, Italy}
\affil[2]{Department of Physics, Faculty of Science, University of Zagreb, 10000 Zagreb, Croatia}
\affil[3]{Horia Hulubei National Institute of Physics and Nuclear Engineering, IFIN-HH, Magurele 077125, Romania}
\affil[4]{Politecnico di Milano, Dipartimento di Elettronica, Informazione e Bioingegneria and INFN Sezione di Milano, 20133 Milano, Italy}
\affil[5]{Stefan-Meyer-Institut für Subatomare Physik, 1090 Vienna, Austria}
\affil[6]{Excellence Cluster Universe, Technische Universität München - Garching, Germany}
\affil[7]{RIKEN - Tokyo, Japan}
\affil[8]{The M. Smoluchowski Institute of Physics, Jagiellonian University, 30-348 Kraków, Poland}
\affil[9]{Museo Storico della Fisica e Centro Studi e Ricerche “Enrico Fermi”, 00184 Roma, Italy}

\begin{document}
	\maketitle
	
	Correspondence: Marco.Miliucci@lnf.infn.it; francesco.sgaramella@lnf.infn.it
	
	\newpage
	\abstract{
		A large-area Silicon Drift Detectors (SDDs) system has been developed by the SIDDHARTA-2 collaboration for high precision light kaonic atoms X-ray spectroscopy at the DA$\Phi$NE collider of Istituto Nazionale di Fisica Nucleare - Laboratori Nazionali di Frascati. The SDDs geometry and electric field configuration, combined with their read-out electronics, make these devices suitable to perform high precision light kaonic atoms spectroscopy measurements in the high background of the DA$\Phi$NE collider. This work presents the spectroscopic response of the SDDs system during the kaonic helium first exotic atoms run of SIDDHARTA-2, preliminary to the kaonic deuterium data taking campaign}
	
	
	\section{Introduction}
	Silicon Drift Detectors (SDDs) technology \cite{1,2} combines the Silicon p-n junction reverse bias properties to an innovative electronic field design, resulting in low electric noise and high-rate capability devices able to perform high precision X-ray measurements for a wide range of applications \cite{3}-\cite{7}. Among these, the SDDs excellent spectroscopic performances are used for light kaonic atoms X-ray spectroscopy to accurately determine the shift ($\epsilon$) and the width ($\Gamma$) of the atomic levels caused by the $\bar{K}$N strong interaction. These measurements allow to probe the non-perturbative Quantum Chromodynamics (QCD) in the strangeness sector, with implications extending from particle and nuclear physics to astrophysics \cite{8}-\cite{11}. In 2009, the SDDs technology has been employed for the first time by the SIDDHARTA collaboration \cite{12}, at the DA$\Phi$NE collider \cite{13,14} of Istituto Nazionale di Fisica Nucleare - Laboratori Nazionali di Frascati (INFN-LNF), achieving the most precise measurement of the K\textsuperscript{-}H fundamental level shift ($\epsilon$) and width ($\Gamma$). Nowadays, the SIDDHARTA-2 collaboration is ready to perform the analogous more challenging kaonic deuterium (K\textsuperscript{-}d) 2p$\rightarrow$1s transition measurement. Monte Carlo simulations \cite{10}, based on theoretical calculations \cite{15,16} and considering one order of magnitude lower yield with respect to the SIDDHARTA K\textsuperscript{-}H measurement \cite{17}, predict a precision of about 30 eV and 80 eV for the extracted shift and width determination, respectively. Furthermore, the Region of Interest (ROI) of the SIDDHARTA-2 goes from 4000 eV to 12000 eV, perfectly matching with the SDDs high quantum efficiency (>85\% for 450 $\mu$m thick Silicon wafer).\\
	To achieve this unprecedented and ambitious goal, the SIDDHARTA-2 collaboration developed a SDDs system dedicated to the kaonic deuterium X-ray spectroscopy measurement. This work presents the SIDDHARTA-2 experimental apparatus and the SDDs system X-ray spectroscopy response during the first phase of the kaonic helium data taking campaign, performed to characterize the apparatus previous to the difficult kaonic deuterium measurement.
	
	\section{The SIDDHARTA-2 experiment}
	The SIDDHARTA-2 experimental apparatus is presently installed at the DA$\Phi$NE collider of INFN-LNF. The cross section layout of the setup is shown in Figure 1 left. The target cell, made by Kapton walls reinforced with aluminium supports, operates at a temperature around 30 K and pressure of 0.4 MPa (equivalent to 3\% LHD), to optimize the kaon stopping efficiency and the X-rays yield. The total X-ray detection active area is 245.8 cm\textsuperscript{2}, given by the 48 SDDs arrays placed around the target cell. The veto system (\cite{18},\cite{19}) is composed of plastic scintillators placed all around the target cell, both external (Veto-1) and internal (Veto-2) to the vacuum chamber, to reject the radiation generated by the nuclear processes within the target cell and that one coming from the machine. The plastic scintillators below and above the beam pipe, working in coincidence mode on the vertical plane (kaon trigger), are used to suppress the electromagnetic background.\\
	The dedicated GEANT4 simulation, taking as input parameters the theoretical calculations \cite{15,16} and an yield of 0.1\% for the $K_{\alpha}$ transition of kaonic deuterium extrapolated from the SIDDHARTA run in 2009 \cite{17}, allows the optimization of each single element of the SIDDHARTA-2 experimental apparatus. The fit resulting from the Monte Carlo simulation, shown in Figure 1 right, estimates that the precision by which both $\epsilon_{1s}$ and $\Gamma_{1s}$ are expected to be measured in a run of 800 pb\textsuperscript{-1} is comparable to the results obtained by the SIDDHARTA for the kaonic hydrogen measurement.\\
	In order to perform the challenging kaonic deuterium high precision X-ray spectroscopic measurement, a new monolithic SDDs system has been developed by Fondazione Bruno Kessler, Politecnico di Milano, INFN-LNF (Italy) and the Stefan Meyer Institute (Austria) in the framework of the SIDDHARTA-2 collaboration. It consists of 450 $\mu$m thick SDDs arrays (2 x 4 matrix with total active area of 5.12 cm$\textsuperscript{2}$, for a active/total surface ratio of 0.75). The Silicon bulk is glued on an alumina carrier providing the polarization to the SDD units and housing the CMOS low-noise charge sensitive preamplifier (CUBE) bonded close to the SDD n$\textsuperscript{+}$ anode. The array is coupled to the dedicated front-end electronics (SFERA) \cite{20,21} for the X-ray signal processing. Each device is screwed on a high purity aluminum support which provides the thermal contact with the target cell frame, to cool them below 170 K, improving their energy and timing resolutions. The structure of the ceramic carrier (gear-wheel type) allows a close packaging of the SDDs arrays around the target cell, optimizing the geometrical efficiency. The SIDDHARTA-2 SDDs system performances have been firstly investigated and optimized in the laboratory \cite{22,23} and then tested in the hard environment of the DA$\Phi$NE collider \cite{24}, revealing good performances in terms of stability, linear conversion, energy and timing resolutions. All these features make these devices suitable to perform high precision kaonic atoms X-ray spectroscopy, even in the high background of the particle collider.\\
	Lastly, a luminosity monitor placed on the horizontal side of the beam pipe (normal with respect to the kaon trigger) continuously monitors the number of kaons and the background generated for e$\textsuperscript{+}$ - e$\textsuperscript{-}$ collisions (see \cite{25} for details). This device is fundamental for the monitoring of the beam quality, specially during the machine optimization phase preliminary to the SIDDHARTA-2 data taking campaign, providing a fast feedback on the beams quality delivered by the collider.\\
	
	\begin{figure}[htbp]
		\centering
		\includegraphics[width=16 cm]{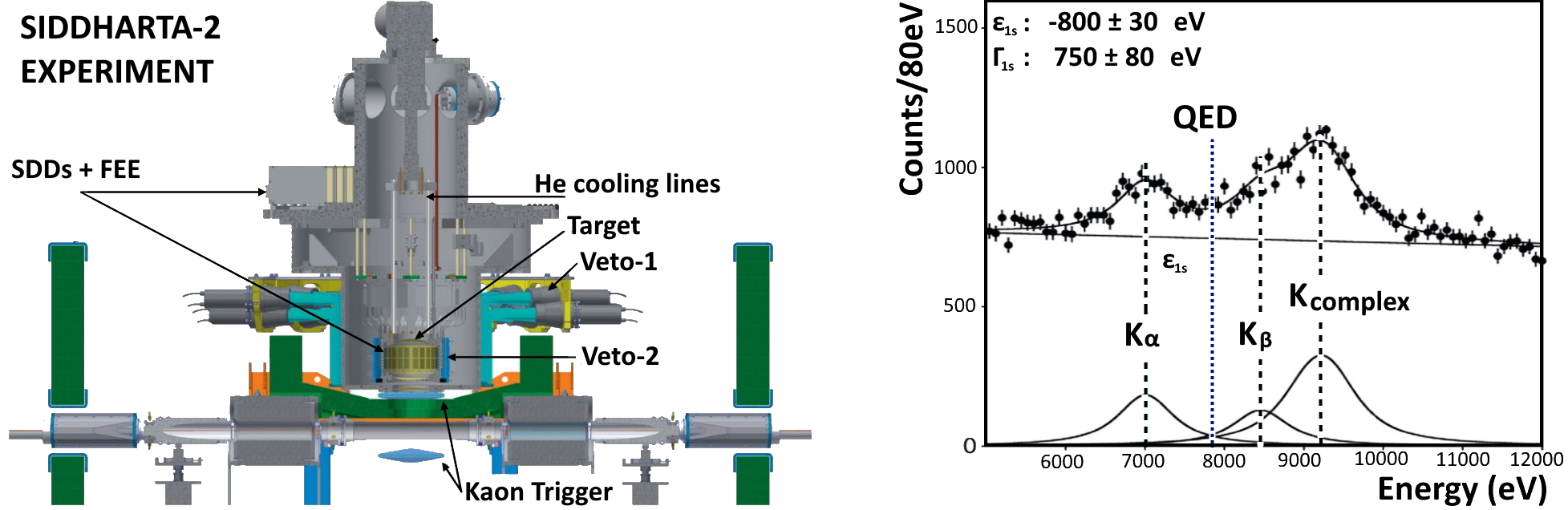}
		\captionsetup{margin=15mm}
		\caption{Left: Cross section layout of the SIDDHARTA-2 experimental apparatus. Right: K\textsuperscript{-}d MC simulation for an integrated luminosity of 800~$pb^{-1}$ considering $\epsilon_{1s}$=-800~eV, $\Gamma_{1s}$=750~eV and yield=0.1\% as input parameters. Elaborated from \cite{10}.\\}
	\end{figure} 
	
	During the DA$\Phi$NE collider e$\textsuperscript{+}$ - e$\textsuperscript{-}$ commissioning, the SIDDHARTA-2 experimental apparatus has been installed in a reduced configuration called SIDDHARTINO, with all SIDDHARTA-2 functionalities but housing a reduced number of SDDs units (8 arrays insted of 48). SIDDHARTINO was running on DA$\Phi$NE from January 2021 to July 2021. After having reached a satisfactory beam quality from the machine in terms of delivered kaons over machine background, the SIDDHARTINO data taking campaign concluded with the K$\textsuperscript{-}$He$\textsuperscript{4}$ L$\textsubscript{$\alpha$}$ X-ray transition measurement.
	
	\section{Results and Discussion}
	The spectroscopic response of the SIDDHARTA-2 SDDs system is evaluated during the K$\textsuperscript{-}$He$\textsuperscript{4}$ data taking campaign, preliminary to the kaonic deuterium measurement. In the Figure 2 plot is shown, as an example, the total SDDs calibrated spectrum corresponding to 10~$pb^{-1}$ of luminosity acquired by the experiment. Thanks to the combination of the linear response at the level few eV with an energy resolution compatible to the common Silicon devices technology  (see \cite{20},\cite{22}), is possible to exactly describe the overall SIDDHARTINO energy response within the SIDDHARTA-2 ROI in terms of background and spurious elements contamination. A simple fit, consisting of two Gaussian functions for the fluorescence peaks and a linear decreasing polynomial for the background, has been used to describe the acquired spectrum. The two main fluorescence peaks are due to the beams products interactions with the components of the experimental apparatus: the Titanium K$\textsubscript{$\alpha$}$ peak at 4509 eV is generated by the the Titanium foil on the top of the target cell, while the Bismuth L$\textsubscript{$\alpha$}$ fluorescence line at 10838 eV is due to the emission from the SDDs ceramics (close to the Silicon wafer).Besides the Ti K$\textsubscript{$\alpha$}$ and Bi L$\textsubscript{$\alpha$}$ peaks, no other invasive spurious elements are detected over a smooth linearly decreasing background.\\
	\begin{figure}[htbp]
		\centering
		\includegraphics[width=16 cm]{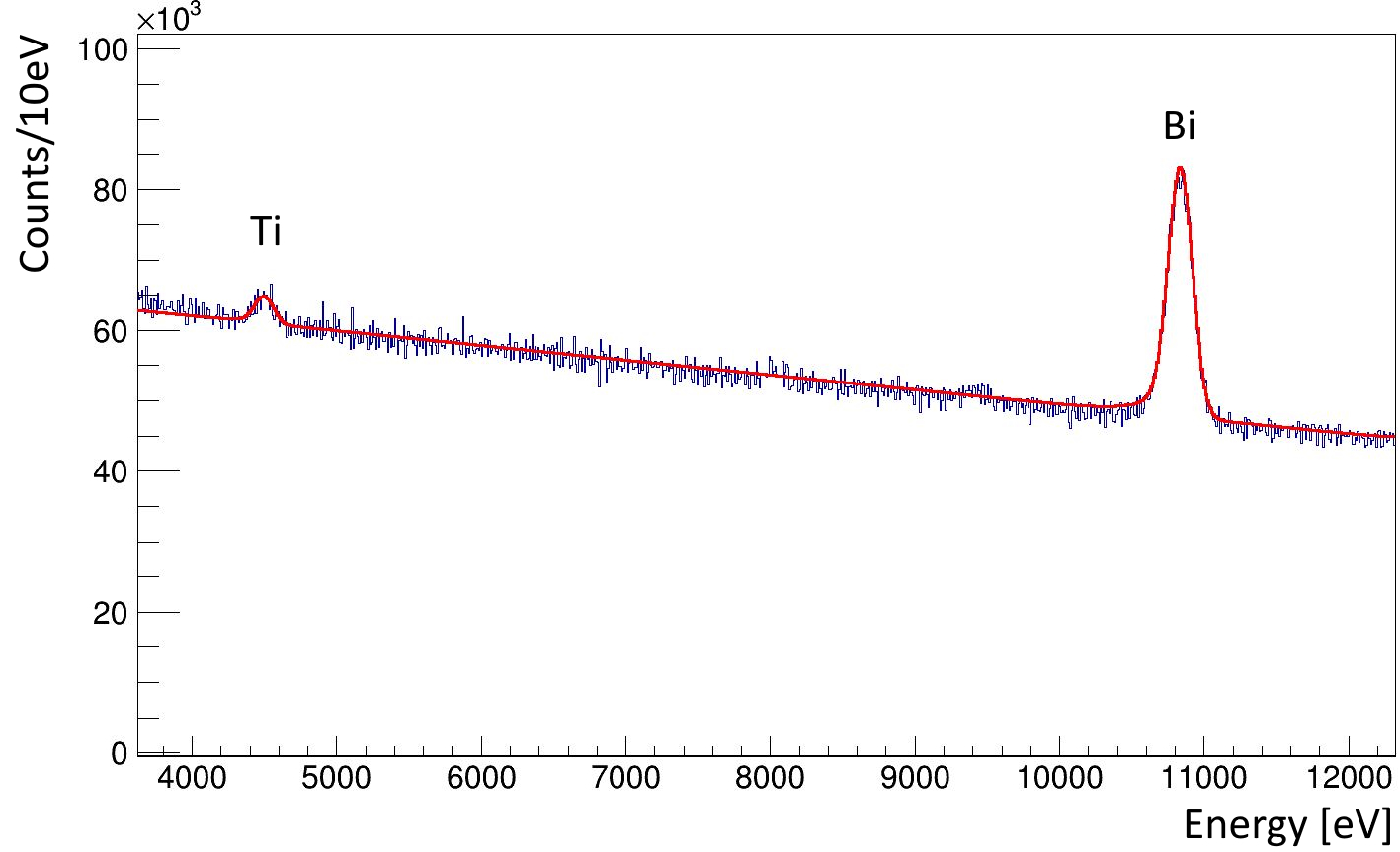}
		\captionsetup{margin=15mm}
		\caption{Total SIDDHARTINO spectrum for 10~$pb^{-1}$ of acquired luminosity. The fit (red) consists of two Gaussian functions for the fluorescence peaks and a first grade polynomial to interpolate the background.}
	\end{figure} 
	
	During the kaonic helium data taking campaign, the Titanium and Bismuth signals have been used for a fast check of the SDDs system linearity.\\
	Figure 3 left shows the 2D-plot of the TDC coincidence signals detected on the scintillators placed respectively on the top (TDC1+TDC2) and the bottom (TDC3+TDC4) of the beam pipe. On the diagonal, we observe a high intensity region corresponding to the kaons distributions (red circle) generated by the decay of the $\Phi$ mesons, well separated from the signals due to the particles lost from e$\textsuperscript{+}$ - e$\textsuperscript{-}$ bunches (MIPs). On the right, the projection in time of the diagonal elements is also presented. The red distribution corresponds to the kaons timing response and it allows to efficiently select the events associated to the kaons. The machine background, in terms of kaons/MIPs, as well as the total luminosity acquired are also evaluated.\\The coincidence signals of the scintillators placed on the vertical side of the beam pipe provides the trigger to the experiment. Thus, it has been also fundamental to exploit the SDDs system timing response in order to disentangle the kaonic helium signal from the asynchronous electromagnetic background of the measurement. The SDDs system timing response, referring to the events associated with a kaon, is shown in Figure 4.\\
	
	\begin{figure}[htbp]
		\centering
		\includegraphics[width=16 cm]{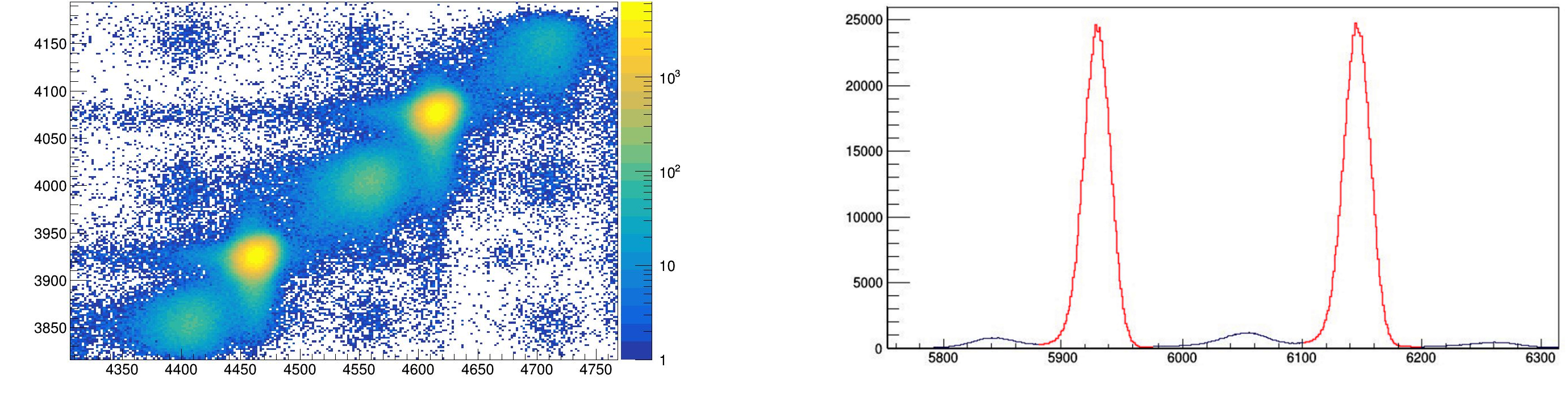}
		\captionsetup{margin=15mm}
		\caption{Kaon trigger distributions. Left: 2D-plot of the TDCs coincidence detected on the top (TDC1 and TDC2) and bottom (TDC3 and TDC4) side of the beam pipe. Right: projection on the time coordinate of the 2D-plot diagonal, red distribution refers to the kaons coincidences.}
	\end{figure} 
	
	\begin{figure}[htbp]
		\centering
		\includegraphics[width=16 cm]{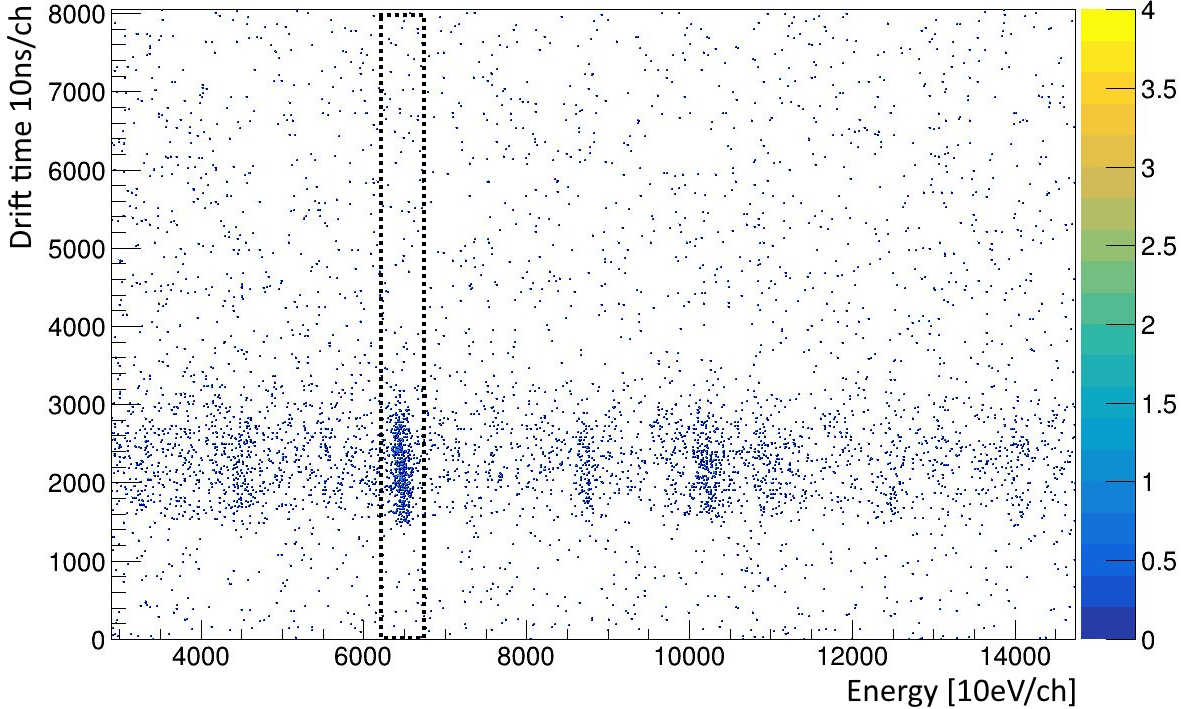}
		\captionsetup{margin=15mm}
		\caption{SIDDHARTA-2 SDDs system timing response. Left: Energy vs drift time 2D-plot given by the coincidence between the signals on the SDDs and kaon trigger for each kaons detected. In between the dotted lines, the energy selection around the K$\textsuperscript{-}$He$\textsuperscript{4}$ L$\textsubscript{$\alpha$}$ peak, from 6200 eV to 6700 eV. Right: SDDs timing distribution of the events with energy within 6200 eV to 6700 eV.\\}
	\end{figure}

	\begin{figure}[htbp]
		\centering
		\includegraphics[width=16 cm]{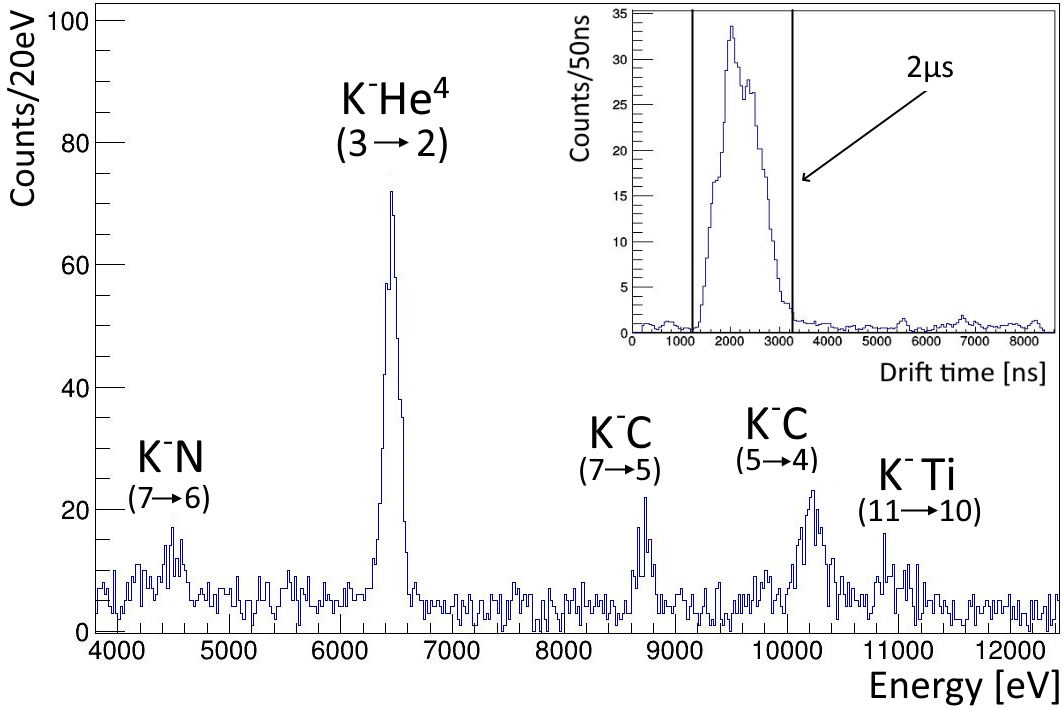}
		\captionsetup{margin=15mm}
		\caption{Calibrated SIDDHARTINO spectrum for 10~$pb^{-1}$ of luminosity acquired after a timing selection procedure. Spectrum obtained applying a 2$\mu$s timing cut of SIDDHARTINO triggered events.}
	\end{figure} 
	
	On the left, the 2D-plot reports the correlation between the energy and timing response for each event detected in the energy range from 4000 eV to 12000 eV for a time interval up to 8 $\mu$s. The kaonic helium signal (distribution confined by the dotted lines around 6400 eV) can be clearly seen within the X-ray events. The histogram on Figure 4 right report the projection on the drift time axis of the kaonic helium events distribution. The plot clearly evidences the timing response of the SDDs system for the selected energy region, allowing to define a cut of 2$\mu$s to reject the asynchronous electromagnetic background.\\
	The calibrated spectrum of triggered events obtained after the timing selection, for the subset of 10~$pb^{-1}$, is shown in Figure 5. The spectrum cleaned by the asynchronous background clearly shows the K$\textsuperscript{-}$He$\textsuperscript{4}$ L$\textsubscript{$\alpha$}$ peak at around 6400 eV and few satellites peaks due to the K$\textsuperscript{-}$ interaction with the elements of the setup. The background of the measurement is smooth and flat, resulting to be reduced by a factor 10$\textsuperscript{-5}$ with respect to the one shown in Figure 2.\\
	Overall, the accurate analysis shown in the present work evaluates for the first time the SDDs system performances during a real light kaonic atom X-ray spectroscopy run at DA$\Phi$NE, proving this technology to be ready to perform the ambitious kaonic deuterium run.      
	
	
	\section{Conclusions}
	A Silicon Drift Detectors (SDDs) system has been developed by the SIDDHARTA-2 collaboration in order to perform the ambitious K\textsuperscript-d 2p $\rightarrow$ 1s transition shift ($\epsilon$) and width ($\Gamma$) measurement. The SIDDHARTA-2 experimental apparatus has been installed at the DA$\Phi$NE collider of INFN-LNF in the reduced (SDDs) configuration, called SIDDHARTINO, during the machine beam commissioning phase concluded by the K$\textsuperscript{-}$He$\textsuperscript{4}$ data taking campaign in July 2021. The spectroscopic response of the SDDs system has been accurately evaluated, proving that the SDDs system is suitable and ready to perform the more challenging K\textsuperscript-d measurement.\\
	The SIDDHARTA-2 data taking campaign is scheduled for 2021-2022. Meanwhile the collaboration is already developing a new 1 mm thick SDDs technology dedicated to the measurement of heavier kaonic atoms to explore higher energy intervals with respect to the SIDDHARTA-2 one, with the aim to obtain additional fundamental information for the non-perturbative QCD in the strangeness sector.
	
	\vspace{6pt} 
	
	Visualization, M.M.; software, M.BR., M.IL., F.N., A.SC. and F.SI.; validation, C.C., C.F., C.G. and J.Z; formal analysis, M.M, L.D.P., A.SC., F.SG., H.S., D.S. and O.V.D.; resources, C.C., C.F., J.Z; investigation, M.M, M.BA., MI.C., A.C., G.D., L.D.P., M.IL., F.N., P.K., A.SC., F.SG., D.S., F.SI. and M.T.; writing---original draft preparation, M.M. and C.C.; writing---review and editing, M.M, D.B., MA.C., R.D.G., M.IW., P.L.S., J.M., P.M., S.N., K.P., M.SI., M.SK. and A.SP.; supervision, C.C., C.F., C.G. and J.Z\\
	
	Part of this work was supported by the Austrian Science Fund (FWF): P24756-N20 and P33037-N; the Croatian Science Foundation under the project IP-2018-01-8570; EU STRONG-2020 project (grant agreement No. 824093), the Polish Ministry of Science and Higher Education grant No. 7150/E-338/M/2018 and the Foundational Questions Institute and Fetzer Franklin Fund, a donor advised fund of Silicon Valley Community Foundation (Grant No. FQXi-RFP-CPW-2008).\\
	
	The authors acknowledge C. Capoccia from INFN-LNF and H. Schneider, L. Stohwasser, and D. Pristauz Telsnigg from Stefan-Meyer-Institut für Subatomare Physik for their fundamental contribution in designing and building the SIDDHARTA-2 setup. We thank as well the DA$\Phi$NE staff for the excellent working conditions and permanent support.\\
	
	The authors declare no conflict of interest.\\

\end{document}